\documentstyle[12pt,twoside,cmp209,epsf]{article}
\title[Phase transitions in binary fluids]{On the theory of phase
transitions in binary fluid mixtures}
\author{O.V.Patsahan}
\address{Institute for Condensed Matter Physics of the National Academy
of Sciences of Ukraine, 1 Svientsitskii Str., UA-290011, Lviv, Ukraine
 }
 
\begin{document}
\maketitle

\begin{abstract}
The microscopic approach to the description of the phase behaviour and
critical phenomena in binary fluid mixtures is proposed. It is based on the
method of collective variables with a reference system. The physical nature
of the order parameter in  a binary mixture is discussed. The  basic density
measure (Ginsburg-Landau-Wilson Hamiltonian) is obtained in the
collective variable phase space which contains the variable connected with
the order parameter of the system. It is shown that the problem can be
reduced to the 3D Ising model in an external field.
\pacs 05.70.Fh, 05.70.Jk, 64.10.+h
\end{abstract}
\renewcommand{\theequation}{\arabic{section}.\arabic{equation}}

\section{Introduction}
\setcounter{equation}{0}
    The study of phase transitions and critical phenomena in multicomponent
fluid systems is very interesting from the theoretical as well as
practical point of view. Whereas in one-component fluid systems only
gas-liquid equilibria exist,three different types of two-phase
equilibria have to be considered in fluid mixtures: gas-liquid,
liquid-liquid and gas-gas equilibria. Despite the numerous
experimental results now available \cite{rowl}- %duren,kumar,schn,
\cite{schn1}, theoretical achievements in understanding a microscopic
mechanism of phase behaviour and nonuniversal critical properties  of
multicomponent fluids are limited. Most of the theoretical studies  devoted
to this problem may be divided  into three main groups: phenomenological
theories, mean field approaches and integral equation methods. The
phenomenological approaches \cite{fisher}- %,grifwh,saam,seng, anisim3,
\cite{anisim4} give predictions about critical exponents and scaling
functions but no quantitative  estimates of nonuniversal  critical
 amplitudes are possible  within this framework. The problem of the
phase  diagram sensitivity  to the microscopic model  also remains
unsolved. The mean field theories \cite{konscot}  and integral
equation methods \cite{wais}-
%,jedr1,jedr2,abramo,monson,caccam,males1,chen, \cite{males3}
reproduce different phase diagram types by varying the microscopic
parameters but give qualitative estimates.

Of special interest  are  Refs.\cite{parol1}-
%,parol2,parol3,
\cite{parol6}
devoted to the study of both the universal and nonuniversal
properties. They are based on the previously proposed approach to the
study of the gas-liquid critical point in a one-component fluid.,
namely, the hierarchical reference theory (HRT) \cite{parol7}. On the
microscopic  Hamiltonian grounds, the HRT develops the renormalization
group structure near a critical point.  Recently this approach was
also used for studying the 3D Ising model \cite{parol8}.

In spite of the doubtless success  of the HRT  the problem remains of
constructing a theory that allows within a unified approach a complete
description of  the phase behaviour of multicomponent continuous
systems beginning with the Hamiltonian and ending with the
thermodynamic functions in the neighbourhood of the phase transition
point. This program has already been accomplished in both the 3D Ising
model \cite{yuk} and a simple fluid near the gas-liquid critical point
\cite{yuk1}-\cite{yukiko}. Within the framework of the $\phi^{4}$ model this
approach has permitted to obtain the non-classical critical exponents and
analytical expressions for thermodynamic functions .

This theory has its origin in the approach based on the functional
representation of a partition function in the collective variables (CV)
space \cite{yukhol}. First the method of CV was proposed for a study
of the charged particles systems\cite{zubar}-\cite{yukhol} and then
it was applied to the second order phase transition theory \cite{yuk}.
The point is that the statistical description of the phase transition
 is to be performed in the appropriate phase space specific for a certain
physical model.  Among the independent variables of this space  there must
be  those connected with order parameters. This phase space forms a set of
CV . Each of them is a mode of density fluctuations corresponding to the
specific  feature of the model under consideration. In particular, for a
magnetic system the CV are variables connected with spin density fluctuation
modes, for a one-component fluid -- with particle density fluctuation modes.
What is the content of the CV for a multicomponent system? We will answer
this question below.

Experiments have shown that fluids and fluid mixtures near
the ordinary critical points belong to the universality class of
Ising-like systems \cite{kumar}.  Thus, a study of critical properties
of multicomponent systems requires, on the one hand, an extension of
the method worked out for one-component fluids and, on the other hand,
their further development. In \cite{patyuk3,patyuk4} we developed
the CV method with a reference system (RS) for the case of a grand
canonical ensemble for a multicomponent system. Within the framework
of this approach the phase diagram of a symmetrical mixture  was
examined in detail \cite{pat}-\cite{Pat}.  Our previous study has
been mainly restricted to the Gaussian model.  In \cite{patyuk4} we
obtained the explicit form of the Ginzburg-Landau-Wilson Hamiltonian
 ($\phi^{4}$ model) for a binary symmetrical  mixture in the vicinity of the
gas-gas demixing critical point.  In this paper we generalize the approach
in the case  of a non-symmetrical binary fluid system (the system of
different size particles  interacting via different potentials).

The layout of the paper is as follows. We give a functional
representation of a grand partition function of a two-component system in
Section 2. The physical nature of the order parameter in a binary mixture is
discussed in Section 3. In Section 4 we construct the basic density measure
(Ginzburg-Landau-Wilson Hamiltonian) with respect to  the CV which
include a variable corresponding to the order parameter.

\section{Functional representation of the grand partition function of
a binary mixture}
\setcounter{equation}{0}

Let us consider a classical two-component system of interacting particles
consisting of $N_{a}$ particles of species $a$ and $N_{b}$ particles of
species $b$. The system is in  volume $V$  at  temperature $T$.

        Let us assume that an  interaction  in  the  system  has a
pairwise additive character. The interaction potential between a
$\gamma$ particle at $\vec r_{i}$  and a $\delta$ particle at $\vec
r_{j}$  may be expressed as a sum  of two terms:
\begin{displaymath}
U_{\gamma\delta}(r_{ij})=\psi_{\gamma\delta}(r_{ij}) +
\phi_{\gamma\delta}(r_{ij}),
\end{displaymath}
where
$\psi_{\gamma\delta}(r)$  is a potential of a short-range repulsion
that will be chosen as an interaction between two hard spheres
$\sigma_{\gamma\gamma}$ and  $\sigma_{\delta\delta}$.
$\phi_{\gamma\delta}(r)$ is an attractive part of the potential which
dominates at large distances. An arbitrary positive
function belonging to the $L_{2}$  class  can  be chosen as the
potential $\phi_{\gamma\delta}(r)$.

     Further consideration of the problem is done in  the  extended
phase space: in the phase space of the Cartesian  coordinates  of  the
particles and in the CV phase space. An interaction connected with
repulsion (potential $\psi_{\gamma\delta}(r)$) is  considered
in  the  space  of the Cartesian coordinates of the particles. We call
this two-component hard spheres system a  reference  system  (RS).
The interaction connected with an attraction (potential
$\phi_{\gamma\delta}(r)$ ) is considered in the CV space. The phase
space overflow is cancelled by introduction of the transition
Jacobian. The contribution of the short-range forces to the long-range
interaction screening is  ensured by averaging this Jacobian over the RS.

     Then a grand partition function  in  the  CV  representation
with a RS can be written as (for details see Appendix~A):
\begin{displaymath}
\Xi=\Xi_{0}\Xi_{1},
\end{displaymath}
where $\Xi_{0}$ is the grand partition function of the RS.
The thermodynamic and structural properties of the RS are assumed to
be known. Although it is known that mixtures with only repulsive interactions
might undergo a phase transition \cite{bibhan}, we assume that in the
region of temperatures, concentrations and densities we are interested
in, thermodynamic functions of the RS remain analytic.
$\Xi_{1}$ has the following form:
\begin{eqnarray}
\Xi_{1} & = & \int
(d\rho)\,(dc)\exp\Big[\beta\mu_{1}^{+}\rho_{0}+\beta\mu_{1}^{-}c_{0}-
\frac{\beta}{2V}\sum_{\vec k}[\tilde V(k)\rho_{\vec k}\rho_{-\vec k}+
\nonumber \\
&&\tilde W(k)c_{\vec k}c_{-\vec k}+\tilde U(k)\rho_{\vec k}c_{-\vec
k}]\Big]J(\rho,c). \label{a2.1}
\end{eqnarray}
Here the following notations are introduced:

$\rho_{\vec k}$ and $c_{\vec k}$ are the CV connected with total
density fluctuation modes and relative density (or concentration)
fluctuation modes in the binary system.

Functions $\mu_{1}^{+}$ and $\mu_{1}^{-}$ have the form:
\begin{equation}
\mu_{1}^{+}=\frac{\sqrt{2}}{2}(\mu_{1}^{a}+\mu_{1}^{b}),\qquad
\mu_{1}^{-}=\frac{\sqrt{2}}{2}(\mu_{1}^{a}-\mu_{1}^{b})
\label{a2.2}
\end{equation}
(the expressions for $\mu_{1}^{\gamma}$ are given in Appendix A) and
are determined from the equations
\begin{eqnarray}
\frac{\partial\ln\Xi_{1}}{\partial\beta\mu_{1}^{+}}& = &\langle
 N\rangle,\\ \label{a2.3a}
\frac{\partial\ln\Xi_{1}}{\partial\beta\mu_{1}^{-}}& = &\langle
 N_{a}\rangle-\langle N_{b}\rangle. \label{a2.3b}
\end{eqnarray}
\begin{eqnarray}
\tilde V(k)&=&(\tilde \phi_{aa}(k)+\tilde \phi_{bb}(k)+2\tilde
\phi_{ab}(k))/2 \nonumber \\
\tilde W(k)&=&(\tilde \phi_{aa}(k)+\tilde \phi_{bb}(k)-2\tilde
\phi_{ab}(k))/2 \nonumber \\
\tilde U(k)&=&(\tilde \phi_{aa}(k)-\tilde \phi_{bb}(k))/2,
\label{a2.4}
\end{eqnarray}
\begin{equation}
J(\rho,c)=\int (d\omega)\,(d\gamma)\exp\Big[i2\pi\sum_{\vec
k}(\omega_{k}\rho_{k}+\gamma_{k}c_{k})\Big]J(\omega,\gamma),\label{a2.5}
\end{equation}
\begin{eqnarray}
J(\omega,\gamma)&=&\exp\Big[\sum_{n\geq 1}\sum_{i_{n}\geq
0}\frac{(-i2\pi)^{n}}{n!}\sum_{\vec k_{1}\ldots\vec
k_{n}}M_{n}^{(i_{n})}(0,\ldots,0)\times\nonumber \\
&  &\gamma_{\vec k_{1}}\ldots\gamma_{\vec
k_{i_{n}}}\omega_{\vec k_{i_{n+1}}}\ldots\omega_{\vec k_{n}}\Big].
\label{a2.6}
\end{eqnarray}
Index $i_{n}$ is used to indicate the number of variables
$\gamma_{\vec k}$ in the cumulant expansion (\ref{a2.5}). Cumulants
$M_{n}^{(i_{n})}$ are expressed as  linear combinations  of the partial
cumulants $M_{\gamma_{1}\ldots\gamma_{n}}$ (see (\ref{dA.3})) and
are presented for $\gamma_{1},\ldots,\gamma_{n}=a,b$  and $n\leq 4$ in
\cite{patyuk4} (see Appendix~B in \cite{patyuk4}).

Formulas (\ref{a2.1})-(\ref{a2.6})  are the initial working formulas
in our study of phase transitions in  binary fluids.

\section{The order parameter in a binary mixture}
\setcounter{equation}{0}
   A choice of the order parameter in multicomponent fluid mixtures is
a serious problem because  the character of the phase transition can
be changing continuously from the pure gas-liquid transition to the
mixing-demixing one. The question of the physical nature of the
order parameter in binary fluid mixtures has been considered until
recently from the point of view of both the phenomenological theory
\cite{anisim3,anisim4} and the microscopic approach \cite{chen}, \cite{parol2},
\cite{patyuk4}, \cite{patyuk1}. Nowdays the
commonly accepted idea is that both the gas-liquid and mixing-demixing phase
transitions are accompanied by total density fluctuations as well as by
relative density (or concentration) fluctuations. This is the only
symmetrical mixture  which exibits a complete distinction between these two
processes \cite{patyuk4}. However, most likely such an "ideal" system does
not occur in reality. In real mixtures the contribution from each type of
the fluctuation  processes changes along the critical curve. The evaluation
of such contributions at each critical curve point is essential to the
definition of the order parameter and to the understanding of the phase
transition character in the mixture. It seems to us that in our approach the
question of the physical nature of the order parameter has a consistent and
clear solution. Here we shall briefly focus on it.

Let us consider a Gaussian approximation of functional integral
(\ref{a2.1})-(\ref{a2.6}). This approximation, also known as the random-phase
approximation, yields the correct qualitative picture of the phenomena under
consideration. As the result of the integration over variables
$\gamma_{k}$ and $\omega_{k}$, $\Xi_{1}$ can be rewritten as %
\begin{eqnarray} \Xi_{1}^{G} &= & \frac{1}{2\pi}{\prod_{\vec
k}}'\frac{1}{\pi}\frac{1}{\sqrt{\Delta(k)}}\int(d\rho)\,(dc)\exp\Big[
\rho_{0}(\beta\mu_{1}^{+}+\aleph_{1}/\Delta)+\nonumber \\
&&c_{0}(\beta\mu_{1}^{-}+\aleph_{2}/\Delta)-(M_{1}^{(0)}\aleph_{1}+M_{1}^{(1)}
\aleph_{2})-\frac{1}{2}\sum_{\vec k}[\rho_{\vec k}\rho_{-\vec
k}\times \nonumber \\
&&A_{11}(k)+
c_{\vec k}c_{-\vec k}A_{22}(k)+2\rho_{\vec k}c_{-\vec k}A_{12}(k)]\Big],
\label{b3.1}
\end{eqnarray}
where
\begin{displaymath}
\aleph_{1}= M_{2}^{(2)}M_{1}^{(0)}-M_{2}^{(1)}M_{1}^{(1)}, \qquad
\aleph_{2}= M_{2}^{(0)}M_{1}^{(0)}-M_{2}^{(1)}M_{1}^{(0)}
\end{displaymath}
\begin{eqnarray}
A_{11}(k)& = & -\frac{1}{2}\Big(\frac{\beta}{V}\tilde
V(k)+\frac{M_{2}^{(2)}}\Delta\Big) \nonumber \\
A_{22}(k)& = & -\frac{1}{2}\Big(\frac{\beta}{V}\tilde
W(k)+\frac{M_{2}^{(0)}}\Delta\Big) \nonumber \\
A_{12}(k)& = &-\frac{1}{2}\Big(\frac{\beta}{V}\tilde
U(k)-\frac{M_{2}^{(1)}}\Delta\Big) \label{b3.2}
\end{eqnarray}
\begin{displaymath}
\Delta=M_{2}^{(0)}M_{2}^{(2)}-(M_{2}^{(1)})^{2}.
\end{displaymath}
In order to determine the phase space of the CV connected with the order
parameters we distinguish independent collective excitations by
diagonalizing the square form in (\ref{b3.1}) by means of the orthogonal
transformation:
\begin{eqnarray}
\rho_{\vec k}& =& A(k)\eta_{\vec k} + B(k)\xi_{\vec k} \nonumber \\
c_{\vec k} & = & C(k)\eta_{\vec k} + D(k)\xi_{\vec k}. \label{b3.3}
\end{eqnarray}
The explicit forms for coefficients $A(k)$, $B(k)$, $C(k)$ and $D(k)$ are
given in Appendix~B.

As a result, we have
\begin{eqnarray}
\Xi_{1}^{G}& = &\frac{1}{2\pi}{\prod_{\vec
k}}'\frac{1}{\pi}\frac{1}{\sqrt{\Delta(k)}}\int(d\eta)\,(d\xi)\exp\Big[
\eta_{0}(AM_{1}+CM_{2})+\nonumber \\
&&\xi_{0}(BM_{1}+DM_{2})-(M_{1}^{(0)}\aleph_{1}+
M_{1}^{(1)}\aleph_{2})/(\Delta(0))-\nonumber\\
&&\frac{1}{2}\sum_{\vec k}(\varepsilon_{11}(k)\eta_{\vec k}\eta_{-\vec k}+
\varepsilon_{22}(k)\xi_{\vec k}\xi_{-\vec k})\Big], \label{b3.4}
\end{eqnarray}
where
\begin{equation}
\varepsilon_{ii}(k)=-(A_{11}(k)+A_{22}(k)\mp\sqrt{(A_{11}(k)-A_{22}(k))^{2}+
4A_{12}^{2}(k)}). \label{b3.5}
\end{equation}

One of the quantities (\ref{b3.5})(or both) tends to zero
as the critical temperature is approached. On the other hand, we have to
find such a CV $\eta_{\vec k^{*}}$  (or $\xi_{\vec k^{*}}$) which is
connected with the order parameter. Index $\vec k^{*}$ must
correspond to the point of minimum of one of the functions
$\varepsilon_{11}(k)$ or $\varepsilon_{22}(k)$ ( or both). These
functions depend on  temperature, attractive potentials $\tilde
\phi_{\gamma\delta}(k)$ and characteristics of the RS. The RS enters
into (\ref{b3.5}) by cumulants $M_{\gamma\delta}(k)$.
$M_{\gamma\delta}(k)$ can be expressed by the Fourier transforms of
the direct correlation functions $C_{\gamma\delta}(k)$ by means of the
Ornstein-Zernike equations for a mixture. In \cite{leb1}
the analytic solution of the Percus-Yevick equation for a hard
sphere binary mixture  was obtained.
%%%%%%%%%%%%%%%%%%%%%%%%%%%%%%%%%%%%%%%%%%%%%%%%%%%%%%%%%%%%%%%%%%%%%%%%
\begin{figure}[htbp]
\begin{center}
\epsfxsize 100mm
\epsfysize 100mm
\epsffile{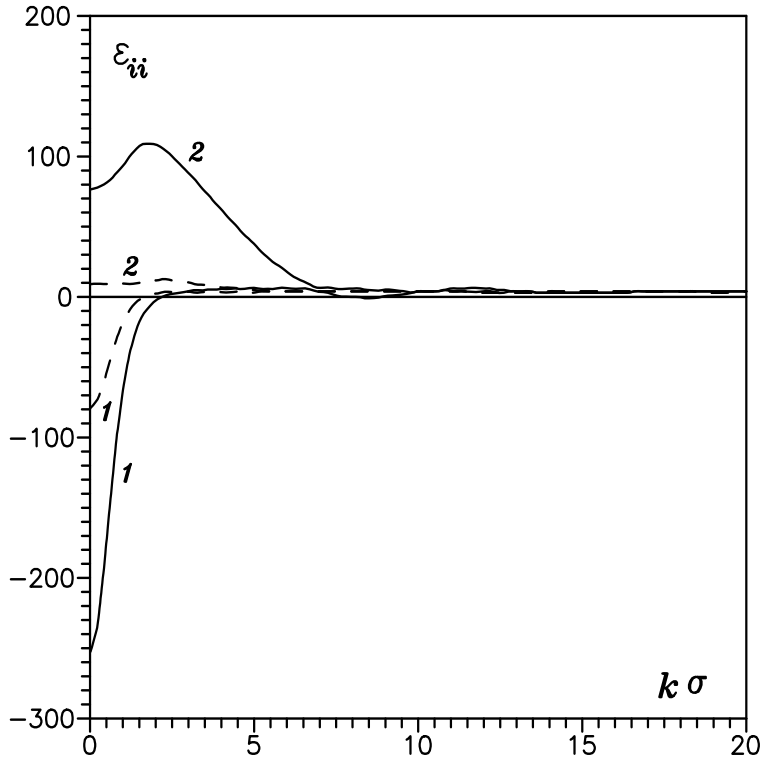}
\end{center}
\caption{Coefficients $\varepsilon_{11}(k)$ and $\varepsilon_{22}(k)$ as
functions of $k$ for the $NH_{3}-N_{2}$ mixture. The solid curves represent
the gas-gas demixing critical point ($T=413^{o}K$, $\eta=0.45$, $x=0.5$)
and the dashed curves represent the gas-liquid critical point
($T=373^{o}K$, $\eta=0.12$, $x=0.5$) \cite{oks1}}
\end{figure}
%%%%%%%%%%%%%%%%%%%%%%%%%%%%%%%%%%%%%%%%%%%%%%%%%%%%%%%%%%%%%%%%%%%%%%%%%

Coefficients $\varepsilon_{11}(k)$ and  $\varepsilon_{22}(k)$ are
studied both as wave vector functions at different values of
temperature $T$, density $\eta$ and concentration $x$ including the
gas-liquid and mixing-demixing critical points (see Fig.~1) \cite{oks1} and
as temperature functions at $\vec k=0$ (see Fig.~2) \cite{Pat}. The results
show that branch $\varepsilon_{11}(k)$ becomes a critical one no matter
whether the system approaches the gas-liquid or gas-gas demixing critical
point. Moreover, $\varepsilon_{11}(k)$ and $\varepsilon_{22}(k)$ have the
minima at $\vec k=0$ \cite{oks1}. Hence we can draw the following
conclusions: \begin{enumerate}
\item Branch $\varepsilon_{11}(k)$ is always critical.

\item Because $\varepsilon_{11}(k)$ has the minimum at $\vec k=0$, the
CV connected with the order parameter is $\eta_{0}$ in the case of the
gas-liquid critical point as well as in the case of the mixing-demixing
phase transition. The particular form of $\eta_{0}$ for each of these
phenomena can be determined by means of the relations between the
microscopic parameters, temperature, density and concentration
of the system, e.g. by means of coefficients $A$, $B$, $C$ and $D$.

\item In the plane $(\rho_{0},c_{0})$ we have distinguished two directions:
the direction of strong fluctuations $\eta_{0}$ and the direction of weak
fluctuations $\xi_{0}$. As a result, we can write the conditions for
the binary mixture critical point in the form:  %
\begin{eqnarray}
\Big[\frac{\partial^{2}\Omega}{\partial\eta_{0}^{2}}\Big]_{c}&=&0,
\label{b3.6a}\\
\Big[\frac{\partial^{2}\Omega}{\partial\eta_{0}\partial\xi_{0}}\Big]_{c}&=&0,
\label{b3.6b}\\
\Big[\frac{\partial^{3}\Omega}{\partial\eta_{0}^{3}}\Big]_{c}&=&0,
\label{b3.6c}
\end{eqnarray}
where $\Omega= -kT\ln\Xi$ is a grand canonical potential.
\end{enumerate}
%%%%%%%%%%%%%%%%%%%%%%%%%%%%%%%%%%%%%%%%%%%%%%%%%%%%%%%%%%%%%%%%%%%%%%%%%
\begin{figure}[htbp]
\begin{center}
\epsfxsize 100mm
\epsfysize 150mm
\epsffile{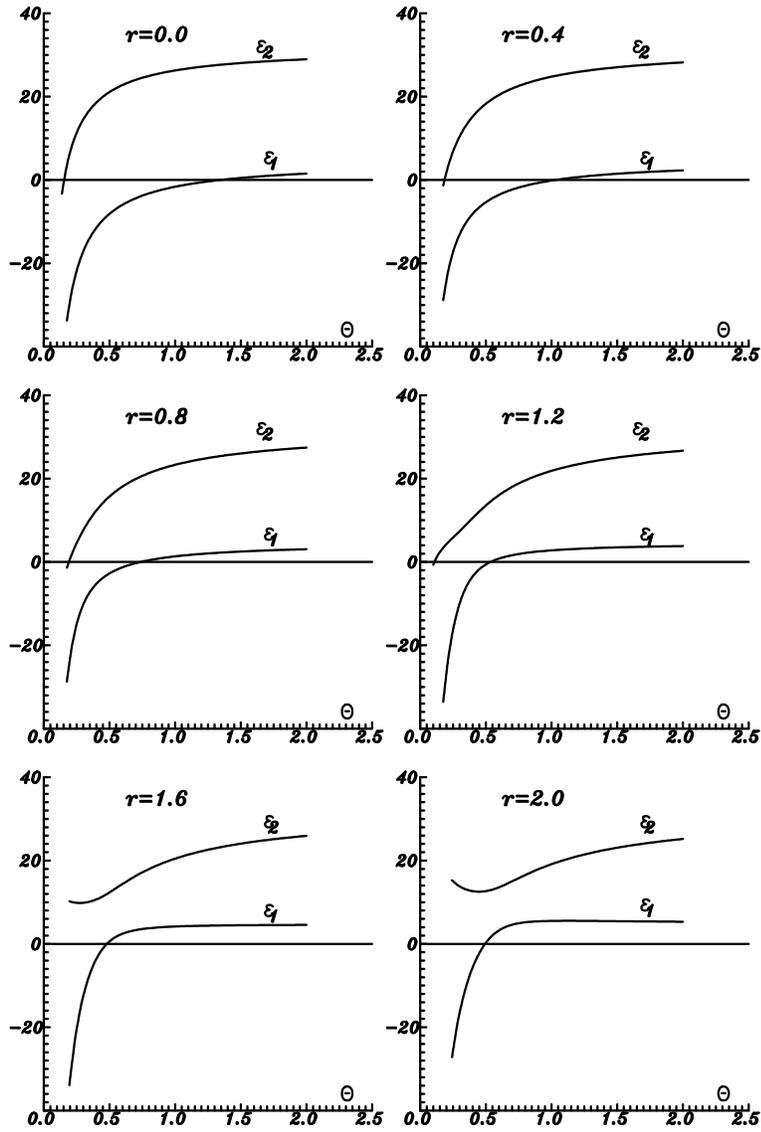}
\end{center}
\caption{Coefficients $\varepsilon_{11}(k=0)$ and $\varepsilon_{22}(k=0)$
as functions of the dimensionless temperature $T$ at $q=1.0$, $\alpha=1.0$,
$\eta=0.26$ and $x=0.7$ for different values of r ($\varepsilon_{ii}(k=0)$
are obtained for a binary hard core Yukawa mixture \cite{oks1}) }
\end{figure}
%%%%%%%%%%%%%%%%%%%%%%%%%%%%%%%%%%%%%%%%%%%%%%%%%%%%%%%%%%%%%%%%%%%%%%%%%

Now let us consider equations (\ref{b3.3}) at $k=0$. From (\ref{b3.3}) it
follows that
\begin{eqnarray}
\eta_{0}& =& \pm D(0)\rho_{0} \mp B(0)c_{0} \nonumber \\
\xi_{0} & = &\mp C(0)\rho_{0}\pm  A(0)c_{0}, \label{b3.7}
\end{eqnarray}
where the upper sign corresponds to the case
\begin{equation}
\vert A_{12}\vert=-A_{12} \qquad (AD-BC=1) \label{b3.8a}
\end{equation}
and the lower sign corresponds to
\begin{equation}
\vert A_{12}\vert=A_{12} \qquad (AD-BC=-1). \label{b3.8b}
\end{equation}
On the other hand, (\ref{b3.7}) can be rewritten as
\begin{eqnarray}
\eta_{0}& =& \pm \rho_{0}\cos \theta \pm c_{0}\sin \theta \nonumber \\
\xi_{0} & = &- \rho_{0}\sin \theta+ c_{0}\cos \theta. \label{b3.9}
\end{eqnarray}
Comparing (\ref{b3.7}) and  (\ref{b3.9}) we can determine rotation
angle $\theta$ of axes $\eta_{0}$ and $\xi_{0}$ in the
$(\rho_{0},c_{0})$ plane from the equation
\begin{equation}
\tan \theta=\frac{C}{A} \label{b3.10}
\end{equation}
(in both cases (\ref{b3.8a}) and  (\ref{b3.8b})). In the case
(\ref{b3.8b}) transformation (\ref{b3.9}) corresponds to both the mirror
image with respect to the $c_{0}$ axis and the rotation in the
$(\rho_{0},c_{0})$ plane. Thus, the proposed approach allows us, on
microscopic grounds, to define the order parameter at each point along a
critical curve and so to understand the phase transition character in the
binary mixture.
%%%%%%%%%%%%%%%%%%%%%%%%%%%%%%%%%%%%%%%%%%%%%%%%%%%%%%%%%%%%%%%%%%%%%%
\begin{figure}[htbp]
\begin{center}
\epsfxsize 100mm
\epsfysize 100mm
\epsffile{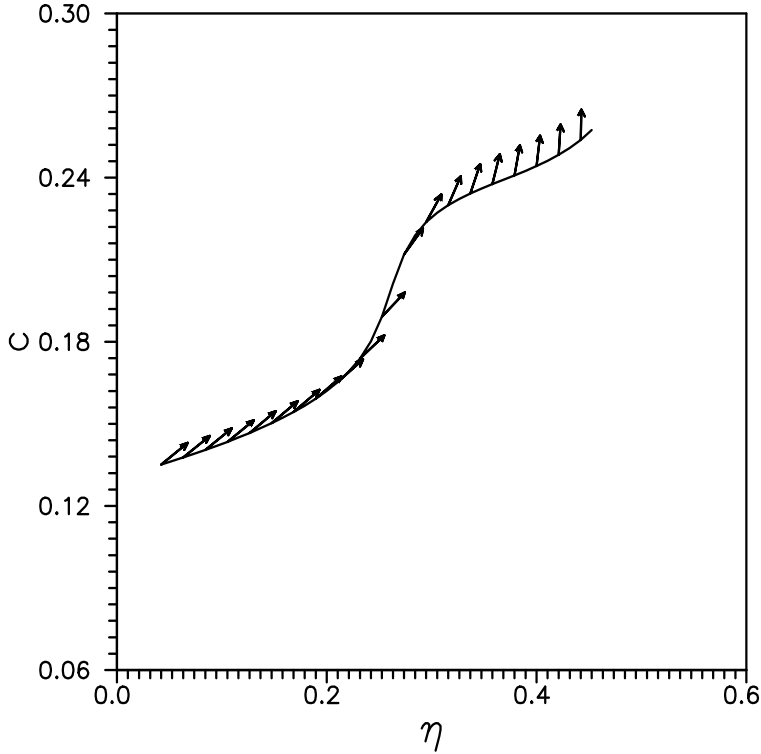}
\end{center}
\caption{Density-concentration projection of the critical line
of the model binary mixture at $\alpha=1.0$, $q=0.9$ and $r=0.6$(mean field
approximation)}
\end{figure}
%%%%%%%%%%%%%%%%%%%%%%%%%%%%%%%%%%%%%%%%%%%%%%%%%%%%%%%%%%%%%%%%%%%%%%%%

Figures 3-5 show the $(\eta,x)$ projections of the $(T,\eta,x)$
critical surfaces of the model binary mixture for various values of
microscopic parameters $\alpha$, $q$ and $r$. The arrows show the direction
of the strong fluctuations (order parameter) along the critical curve in
accordance with formula (\ref{b3.10}). Here the following notations are
introduced: $\eta$ is the packing density ($\eta=\eta_{a}+\eta_{b}$,
$\eta_{i}=\pi\rho_{i}\sigma_{ii}^{3}/6$, $\rho_{i}=\langle N\rangle/V$ is
the number density of species $i$), $x$ is the concentration
($x=\langle N_{b}\rangle/\langle N\rangle$),
$\alpha=\sigma_{aa}/\sigma_{bb}$ is the hard sphere ratio, $\sigma_{ii}$ is
the hard sphere diameter,$q=-\tilde \phi_{bb}(0)/\mid \tilde
\phi_{aa}(0)\mid$  is the dimensionless "like" interaction strength and
$r=-\tilde \phi_{ab}(0)/\mid \tilde \phi_{aa}(0)\mid$ is the  "unlike"
interaction strength (the form of $\phi_{ij}(r)$ is not specified). The
critical surface is derived from the $(f_{mf},V,x)$ surface by the equations
for a binary mixture critical point (in terms of the Helmholtz free energy
\cite{rowl}). $f_{mf}$ is the Helmholtz free energy of a binary mixture in
the mean field approximation (see Appendix C).
% %%%%%%%%%%%%%%%%%%%%%%%%%%%%%%%%%%%%%%%%%%%%%%%%%%%%%%%%%%%%%%%%%%%%%%%%%%
\begin{figure}[htbp]
\begin{center}
\epsfxsize 100mm
\epsfysize 100mm
\epsffile{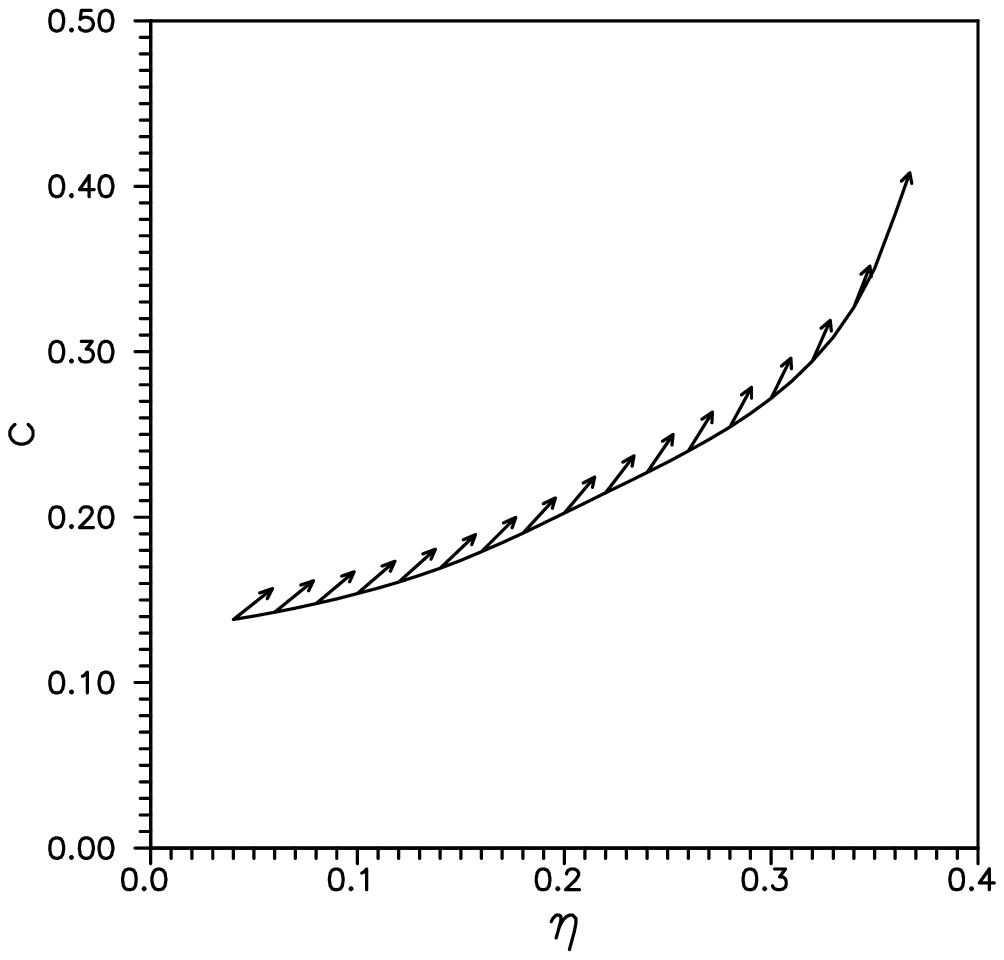}
\end{center}
\caption{Same as figure 3 at $\alpha=0.9$, $q=0.9$ and
$r=0.6$}
\end{figure}
%%%%%%%%%%%%%%%%%%%%%%%%%%%%%%%%%%%%%%%%%%%%%%%%%%%%%%%%%%%%%%%%%%%%%%%%%%%

The purpose of our further study is the calculation of the binary mixture
behaviour in the vicinity of its critical points. Based on the Gaussian
distribution (\ref{b3.1})-(\ref{b3.2}) we have determined the critical
branch and, correspondingly, CV $\eta_{0}$ connected with the order parameter.
Now we shall construct the basic density measure with respect to CV $\eta_{\vec
k}$. As it is shown in \cite{yuk}, in the vicinity of the critical
point the basic density measure exists which includes higher powers of
CV than the second power. We shall follow the program: (1) having
passed from CV $\rho_{\vec k}$ and $c_{\vec k}$ to CV $\eta_{\vec
k}$ and $\xi_{\vec k}$ in (\ref{a2.1}), we shall integrate over
variables $\xi_{\vec k}$ with the Gaussian density measure; (2) then
we shall construct the basic density measure with respect to variables
$\eta_{\vec k}$ (Ginzburg-Landau-Wilson Hamiltonian). We shall
restrict our consideration to the $\eta^{4}$ model.
%%%%%%%%%%%%%%%%%%%%%%%%%%%%%%%%%%%%%%%%%%%%%%%%%%%%%%%%%%%%%%%%%%%%%%%%%
\begin{figure}[htbp]
\begin{center}
\epsfxsize 100mm
\epsfysize 100mm
\epsffile{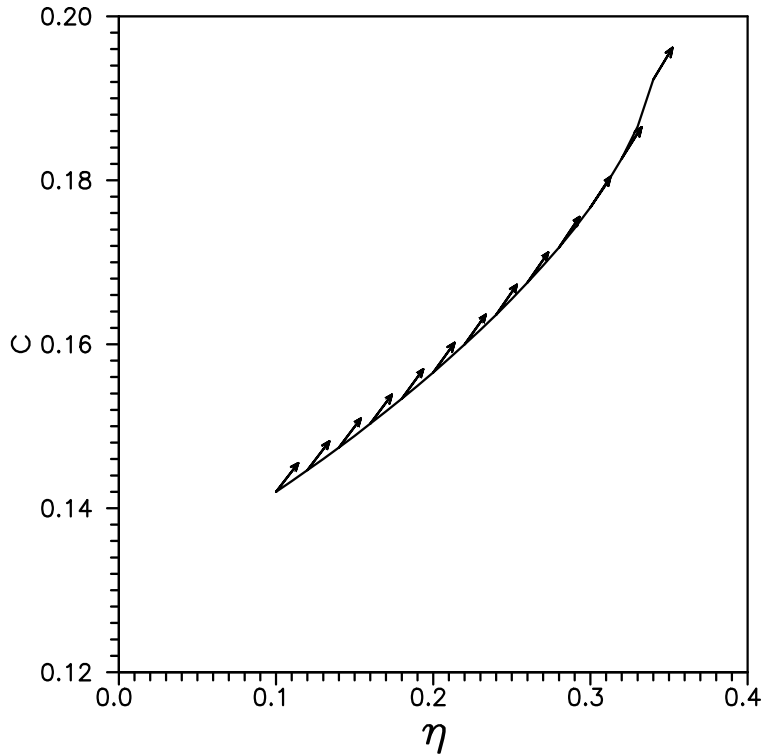}
\end{center}
\caption{Same as figure 3 at $\alpha=0.9$, $q=0.9$ and
$r=0.8$}
\end{figure}
%%%%%%%%%%%%%%%%%%%%%%%%%%%%%%%%%%%%%%%%%%%%%%%%%%%%%%%%%%%%%%%%%%%%%%%%%%%%
\section{The microscopic Ginzburg-Landau-Wilson Hamiltonian in the vicinity
of the binary mixture critical point}
\setcounter{equation}{0}
We pass in (\ref{a2.5}) from CV $\rho_{\vec k}$ and $c_{\vec k}$ to CV
$\eta_{\vec k}$ and $\xi_{\vec k}$:
\begin{eqnarray}
\Xi_{1} & = & \int(d\eta)\,(d\xi)\exp\Big[\eta_{0}\tilde
\mu_{1}+\xi_{0}\tilde \mu_{2}-\frac{1}{2}\sum_{\vec k}[\eta_{\vec
k}\eta_{-\vec k}P(k)+ \nonumber \\
&  & \xi_{\vec k}\xi_{-\vec k}R(k)+2\xi_{\vec k}\eta_{-\vec
k}Q(k)]\Big]J(\eta,\xi), \label{c4.1}
\end{eqnarray}
where
\begin{equation}
\tilde \mu_{1}=\beta(A\mu_{1}^{+}+C\mu_{1}^{-}), \qquad
\tilde \mu_{2}=\beta(B\mu_{1}^{+}+D\mu_{1}^{-})\label{c4.2}
\end{equation}
\begin{eqnarray}
P(k) & = & \frac{\beta}{V}(A^{2}\tilde V(k) +C^{2}\tilde W(k)+2AC\tilde U(k))
\label{c4.3a} \\
R(k) & = & \frac{\beta}{V}(B^{2}\tilde V(k) +D^{2}\tilde W(k)+2BD\tilde U(k))
\label{c4.3b}\\
Q(k) & = & \frac{\beta}{V}(AB\tilde V(k) +CD\tilde W(k)+(AD+BC)\tilde
U(k))
\label{c4.3c}
\end{eqnarray}
\begin{equation}
J(\eta,\xi) = \int(d\chi)\,(d\vartheta)\exp\Big[i2\pi\sum_{\vec
k}(\eta_{\vec k}\chi_{\vec k}+\xi_{\vec k}\vartheta_{\vec k})+
\sum_{n\geq 1}\bar D_{n}(\chi,\vartheta) \label{c4.4}
\end{equation}
\begin{eqnarray}
\bar D_{n}(\chi,\vartheta)&=&\frac{(-i2\pi)^{n}}{n!}\sum_{\vec
k_{1}\ldots\vec k_{n}}\bar M_{n}^{(i_{n})}(0,\ldots)\times \nonumber \\
&& \vartheta_{\vec k_{1}}\vartheta_{\vec k_{2}}\ldots\vartheta_{\vec
k_{n-i_{n}}}\chi_{\vec k_{n-(i_{n+1})}}\ldots\chi_{\vec k_{n}} \label{c4.5}
\end{eqnarray}
\begin{displaymath}
\chi_{\vec k}= A\omega_{\vec k}+C\gamma_{\vec k}, \qquad
\vartheta_{\vec k}= B\omega_{\vec k}+D\gamma_{\vec k}
\end{displaymath}
and $\bar M_{n}^{(i_{n})}$ are linear combinations of cumulants
$M_{n}^{(i_{n})}$.

The square form in (\ref{c4.1}) is diagonal if
\begin{equation}
Q\equiv 0.\label{c4.6}
\end{equation}
Taking into account formulas from Appendix B, equation (\ref{c4.6})
can be rewritten in the form:
\begin{equation}
\frac{(1-x)S_{aa}-xS_{bb}}{\sqrt{x(1-x)}S_{ab}}-\frac{q-1}{r}=0,
\label{c4.7}
\end{equation}
where $S_{ij}(k)$ is a two-particle partial structure factor of the RS.

On the other hand, the square form in (\ref{c4.5}) is diagonal if
\begin{equation}
\bar M_{2}^{(1)}\equiv 0 \label{c4.9}
\end{equation}
and the last equality holds if (\ref{c4.6}) holds. Equation (\ref{c4.6})
holds in the following cases:
\begin{itemize}
\item  a symmetrical mixture: $S_{aa}=S_{bb}$, $x=1/2$, $q=1$

\item a non-symmetrical mixture:$(1-x)S_{aa}-xS_{bb}=0$, $q=1$

\item a non-symmetrical mixture: $q \not=1$.
\end{itemize}

Let us eliminate the linear term in (\ref{c4.5}) by the shift
\begin{displaymath}
\xi_{\vec k}=\xi_{\vec k}^{'}+\bar M_{1}^{(0)}\delta_{\vec k}
\end{displaymath}
and present $\bar D_{n}(\chi,\vartheta)$ as a sum of two terms:
\begin{displaymath}
\bar D_{n}(\chi,\vartheta)= \bar D_{n}^{'}+ \bar D_{n}^{''},
\end{displaymath}
where
\begin{displaymath}
\bar D_{n}^{'}=\bar D_{n}(\vartheta)+\bar D_{n}(\chi,\vartheta)    , \qquad
\bar D_{n}^{''}=\bar D_{n}(\chi).
\end{displaymath}
Here $D_{n}(\vartheta)$ includes only the products of variables
$\vartheta_{\vec k}$,$\bar D_{n}(\chi,\vartheta)$ includes the mixed
products of both variables $\vartheta_{\vec k}$ and $\chi_{\vec k}$,
$\bar D_{n}(\chi)$ includes only the products of variables $\chi_{\vec
k}$. Let us consider the integral
\begin{eqnarray}
{\cal{I}} &= & \int
(d\xi)\exp\Big[{\cal {M}}^{+}\xi_{0}-\frac{1}{2}\sum_{\vec k}\xi_{\vec
k}\xi_{-\vec k}R(k)+i2\pi\sum_{\vec k}\xi_{\vec k}\vartheta_{\vec k}+
\nonumber \\
&& \frac{(-i2\pi)^{2}}{2!}\sum_{\vec k}\bar
M_{2}^{(0)}\vartheta_{\vec k}\vartheta_{-\vec k}\Big]\Big[1+{\cal{A}}
+\frac{1}{2}{\cal{A}}^{2} +\ldots\Big], \label{c4.10}
\end{eqnarray}
where the following notations are introduced:
\begin{equation}
{\cal M^{+}}= \tilde \mu_{2}-R(0)\bar M_{1}^{(0)},
\qquad {\cal A}=\sum_{n\geq 3}\bar D_{n}^{'}, \label{c4.11}
\end{equation}
(in (\ref{c4.10}) the prime on $\xi_{\vec k}$ is omitted
for clarity).

If operator $\frac{\partial}{\partial\xi_{\vec k}}$ is substituted for
$i2\pi\vartheta_{\vec k}$, (\ref{c4.11}) can be rewritten as
\begin{displaymath}
{\cal I}=\Xi_{\xi}^{G}(1+\langle\hat {\cal A}\rangle+\frac{1}{2}\langle\hat
{\cal A}^{2}\rangle+\ldots),
\end{displaymath}
where
\begin{eqnarray}
\langle\ldots\rangle & =& \frac{1}{\Xi_{\xi}^{G}}\frac{1}{2}\prod_{\vec k}
(\pi\bar M_{2}^{(0)}(k))^{-\frac{1}{2}}\exp({\cal
F})\int(d\xi)\exp\Big[{\cal M^{+}}\xi_{0}- \nonumber \\
&& \frac{1}{2}\sum_{\vec k}\xi_{\vec k}\xi_{-\vec
k}R(k)\Big]\ldots\exp\Big[-\frac{1}{2}\sum_{\vec k}\frac{\xi_{\vec
k}\xi_{-\vec k}}{\bar M_{2}^{(0)}}\Big] \nonumber
\end{eqnarray}
\begin{equation}
\Xi_{\xi}^{G}= exp({\cal F})\prod_{\vec k}\sqrt{R(k)\bar
M_{2}^{(0)}(k)+1}\exp\left(\frac{(\bar M^{+})^{2}\bar M_{2}^{(0)}}{2(R(0)\bar
M_{2}^{(0)}+1)}\right), \label{c4.11a}
\end{equation}
\begin{equation}
{\cal F}=\bar M_{1}^{(0)}({\cal M}^{+}+\frac{1}{2}R(0)\bar M_{1}^{(0)})
\label{c4.12}
\end{equation}
Finally, after the integration in (\ref{c4.1}) over variables
$\xi_{\vec k}$ we obtain
\begin{equation}
\Xi=\Xi_{0}\Xi_{\xi}^{G}\Delta^{\xi}\int(d\eta)\exp\Big[\tilde
\mu_{1}\eta_{0}-
\frac{1}{2}\sum_{\vec k}\eta_{\vec k}\eta_{-\vec k}P(k)\Big]J(\eta),
\label{c4.13}
\end{equation}
where $\Delta^{\xi}$ is the result of integrating over $\xi_{\vec k}$ which
does not include $\chi_{\vec k}$. $J(\eta)$ has the form:
\begin{eqnarray}
J(\eta) & = & \int(d\chi)\exp\Big[i2\pi\sum_{\vec k}\chi_{\vec k}\eta_{\vec
k}-i2\pi M_{1}(0)\sum_{\vec k}\chi_{\vec k}\delta_{\vec
k}+\frac{(-i2\pi)^{2}}{2!}M_{2}(0)\sum_{\vec k}\chi_{\vec k}\chi_{-\vec k}+
\nonumber \\
&& \frac{(-i2\pi)^{3}}{3!}M_{3}(0,\ldots)\sum_{\vec k_{1}\vec
k_{2}\vec k_{3}}\chi_{\vec k_{1}}\chi_{\vec k_{2}}\chi_{\vec
k_{3}}\delta_{\vec k_{1}+\vec k_{2} +\vec
k_{3}}+\frac{(-i2\pi)^{4}}{4!}M_{4}(0,\ldots)\times \nonumber \\
&&\sum_{\vec k_{1}\vec k_{2}\vec
k_{3}\vec k_{4}}\chi_{\vec k_{1}}\chi_{\vec k_{2}}\chi_{\vec
k_{3}}\chi_{\vec k_{4}}\delta_{\vec k_{1}+\vec k_{2} +\vec k_{3}+\vec
k_{4}\Big]. \label{c4.14}}
\end{eqnarray}
Here
$$
M_{n}(0,\ldots)=\bar M_{n}^{(n)}(0,\ldots)+\triangle M_{n}(0,\ldots).
$$
Setting ${\cal M}^{+}=0$ we get the following expressions for
$\triangle M_{n}(0,\ldots)$:
\begin{equation}
\triangle M_{1}=\frac{\bar M_{3}^{(1)}}{2\langle N\rangle}\sum_{\vec
k^{'}}\tilde g(k^{'})+\ldots \label{c4.15}
\end{equation}
\begin{eqnarray}
\triangle M_{2}&=&\frac{\bar M_{4}^{(2)}}{2!\langle N\rangle}\sum_{\vec
k^{'}}\tilde g(k^{'})+\frac{(\bar M_{3}^{(1)})^{2}}{2\langle N\rangle^{2}}
\sum_{\vec k^{'}}\tilde g(k^{'})\tilde g(\vert\vec k-\vec k^{'}\vert)+
\nonumber \\
&& \frac{(\bar M_{4}^{(1)})^{2}}{3!\langle N\rangle^{3}}
\sum_{\vec k^{'}\vec k^{''}}\tilde g(k^{'})\tilde g(k^{''})
\tilde g(\vert\vec k-\vec k^{'}-\vec k^{''}\vert)+\ldots \label{c4.15b}
\end{eqnarray}
\begin{eqnarray}
\triangle M_{3}&=&\frac{3}{2}\bar M_{3}^{(2)}\bar
M_{4}^{(1)}\frac{1}{\langle N\rangle}\tilde g(k_{2})\frac{1}{\langle
N\rangle}\sum_{\vec k^{'}}\tilde g(k^{'})+\frac{3}{2}\bar
M_{3}^{(1)}\bar M_{4}^{(2)}\times \nonumber \\
&&\frac{1}{\langle N\rangle^{2}}\sum_{\vec k^{'}}\tilde g(k^{'})\tilde
g(\vert\vec k_{2}+\vec k_{3}-\vec k^{'}\vert)+\ldots\label{c4.15c}
\end{eqnarray}
\begin{eqnarray}
\triangle M_{4}& = &3(\bar M_{3}^{(2)})^{2}\frac{1}{\langle
N\rangle}\tilde g(\vert\vec k_{3}+\vec k_{4}\vert)+2\bar M_{4}^{(1)}\bar
M_{4}^{(3)}
\frac{1}{\langle N\rangle^{2}}\times \nonumber \\
&&\sum_{\vec k^{'}}\tilde
g(k^{'})\tilde g(\vert\vec k_{2}+\vec k_{3}+\vec k_{4}\vert)+\frac{3}{2}
(\bar M_{4}^{(2)})^{2}
\frac{1}{\langle N\rangle^{2}}\times \nonumber \\
&&\sum_{\vec k^{'}}\tilde g(k^{'})\tilde
g(\vert\vec k_{3}+\vec k_{4}-\vec k^{'}\vert)+\ldots,\label{c4.15d}
\end{eqnarray}
where
$$
\tilde g(k)=-\frac{\tilde R(k)}{1+\tilde R(k)\bar S_{2}^{(0)}}
$$
\begin{displaymath}
\tilde R(k)=R(k)\langle N\rangle, \qquad  \bar S_{2}^{(0)}=\bar
M_{2}^{(0)}/\langle N\rangle.
\end{displaymath}
An estimation of corrections $\triangle M_{n}$ was carried out for the
symmetrical mixture in \cite{patyuk4}. In this case
$$
M_{2}=1+\delta_{1},\qquad M_{3}=0, \qquad M_{4}=-2+\delta_{2},
$$
where $\delta_{i}$ are small values.

Let us consider formula (\ref{c4.3a}) for $P(k)$. Substituting into
(\ref{c4.3a})  coefficients $A$, $B$,$C$ and $D$ from Appendix B, we get
\begin{eqnarray}
P(k)&=& \frac{\beta}{V}\Big[\frac{1}{2}(\tilde \phi_{aa}(k)+\tilde
\phi_{bb}(k))+\frac{1}{\sqrt{4A_{12}^{2}+(A_{11}-A_{22})^{2}}}\times \nonumber \\
&&[\tilde \phi_{ab}(k)
(A_{11}-A_{22})+(\tilde \phi_{aa}(k)-\tilde
\phi_{bb}(k))A_{12}]\Big], \label{c4.16}
\end{eqnarray}
where $A_{11}$, $A_{22}$ and $A_{12}$ are functions of temperature.
Using condition (\ref{c4.6}), $P(k)$ can be represented as
\begin{eqnarray}
P(k)\vert_{Q=0}&=&\frac{\beta}{2V}\Big[\tilde \phi_{aa}(k)+\tilde
\phi_{bb}(k) +2\frac{A_{11}-A_{22}}{\vert
A_{11}-A_{22}\vert}\frac{1}{\sqrt{1+\kappa^{2}}}\times \nonumber \\
&&(\tilde \phi_{ab}(k)+
\kappa(\tilde \phi_{aa}(k)-\tilde \phi_{bb}(k))\Big], \label{c4.17}
\end{eqnarray}
where
$$
\kappa=\frac{q-1}{r}=\frac{(1-x)S_{aa}-xS_{bb}}{2\sqrt{x(1-x)}S_{ab}}.
$$
$P(k)$ takes both negative ( at small $\vert \vec k\vert$)  and
positive ( at large $\vert \vec k\vert$) values . In the region
$\vert \vec k\vert>B$ (see Fig.~6) we can integrate in (\ref{c4.13}) over
$\chi_{\vec k}$ and $\eta_{\vec k}$ with the Gaussian measure density
as the basic one. As a result, we get in the approximation of the $\eta^{4}$
model
\begin{eqnarray}
\Xi& = & \Xi_{0}\Xi_{\xi}^{G}\Delta^{\xi}\Delta^{\eta}\int d(\eta)^{N_{B}}\,
d(\chi)^{N_{B}}\exp\Big[\tilde \mu_{1}\eta_{0}-\frac{1}{2}\sum_{\vec k}
P(k)\eta_{\vec k}\eta_{-\vec k}+ \nonumber \\
&&i2\pi\sum_{\vec k}\chi_{\vec k}\eta_{\vec
k}-i2\pi \bar M_{1}(0)\sum_{\vec k}\chi_{\vec k}\delta_{\vec
k}+\frac{(-i2\pi)^{2}}{2!}\bar M_{2}(0)\sum_{\vec k}\chi_{\vec
k}\chi_{-\vec k}+ \nonumber \\
&& \frac{(-i2\pi)^{3}}{3!}\bar M_{3}(0,\ldots)\sum_{\vec k_{1}\vec
k_{2}\vec k_{3}}\chi_{\vec k_{1}}\chi_{\vec k_{2}}\chi_{\vec
k_{3}}\delta_{\vec k_{1}+\vec k_{2} +\vec
k_{3}}+\frac{(-i2\pi)^{4}}{4!}\bar M_{4}(0,\ldots)\times \nonumber \\
&&\sum_{\vec k_{1}\vec k_{2}\vec
k_{3}\vec k_{4}}\chi_{\vec k_{1}}\chi_{\vec k_{2}}\chi_{\vec
k_{3}}\chi_{\vec k_{4}}\delta_{\vec k_{1}+\vec k_{2} +\vec k_{3}+\vec
k_{4}\Big], \qquad   \vert\vec k_{i}\vert<B, \label{c4.18}}
\end{eqnarray}
where in place of $M_{i}(0,\ldots)$ stand renormalized coefficients
$\bar M_{i}(0,\ldots)$:
$$
\bar M_{i}(0,\ldots)=M_{i}(0,\ldots)+\triangle\bar M_{i}(0,\ldots)
$$
\begin{displaymath}
\Delta^{\eta}=\prod_{\vert\vec k\vert>B}\sqrt{P(k)\bar M_{2}+1}
\end{displaymath}
We can consider a set of $\vec k$ vectors, $\vert\vec k\vert<B$, as
corresponding to the sites of a reciprocal lattice conjugated to a
certain block lattice ${r_{l}}$ with $N_{B}$ block sites in  volume
$V$:
$$
\langle N_{B}\rangle=V(B/\pi)^{3}=\frac{\langle
N\rangle(B\sigma_{bb})^{3}(x+\alpha^{3}(1-x))}{6\pi^{2}\eta}.
$$
One may consider quantity $B$ as the size of the first Brillouin zone of
this block lattice.

Next, two shifts are carried out in order to eliminate the cubic term in
(\ref{c4.18}) \cite {yuk2}:
$$
\chi_{\vec k}=\chi_{\vec k}^{'} +\Delta\delta_{\vec k}, \qquad
\eta_{\vec k}=\eta_{\vec k}^{'} +\tilde M_{1}\delta_{\vec k},
$$
where
\[
\Delta = -i2\pi\bar M_{3}
\]
and
\[
\tilde M_{1} = \bar M_{1} -\frac{\bar M_{2}\bar M_{3}}{\bar M_{4}} +
\frac{1}{3}\frac{(\bar M_{3})^{3}}{(\bar
M_{4})^{2}}.
\]
Then, (\ref{c4.18}) has the form:
\begin{eqnarray}
\Xi& = & \Xi_{0}\Xi_{\xi}^{G}\Delta^{\xi}\Delta^{\eta}\exp({\cal G})\int
d(\eta^{'})^{N_{B}}\, d(\chi^{'})^{N_{B}}\exp\Big[
h\eta_{0}^{'}-\frac{1}{2}\sum_{\vec k} P(k)\times \nonumber \\
&&\eta_{\vec k}^{'}\eta_{-\vec
k}^{'}+
i2\pi\sum_{\vec k}\chi_{\vec k}^{'}\eta_{\vec
k}^{'}+\frac{(-i2\pi)^{2}}{2!}\tilde M_{2}(0)\sum_{\vec k}\chi_{\vec
k}^{'}\chi_{-\vec k}^{'}+
\frac{(-i2\pi)^{4}}{4!}\times \nonumber \\
&&\tilde M_{4}(0,\ldots)
\sum_{\vec k_{1}\vec k_{2}\vec
k_{3}\vec k_{4}}\chi_{\vec k_{1}}^{'}\chi_{\vec k_{2}}^{'}\chi_{\vec
k_{3}}^{'}\chi_{\vec k_{4}}^{'}\delta_{\vec k_{1}+\vec k_{2} +\vec
k_{3}+\vec k_{4}\Big], \qquad   \vert\vec k_{i}\vert<B, \label{c4.19}}
\end{eqnarray}
where
\[
\tilde M_{2}(0)=\bar M_{2}(0)-\frac{1}{2}\frac{(\bar M_{3}(0))^{2}}{\bar
M_{4}(0)}, \qquad \tilde M_{4}(0,\ldots)=\bar M_{4}(0,\ldots),
\]
\[
{\cal G}=-\frac{\bar M_{3}}{\bar M_{4}}\left(\bar M_{1}-\frac{\bar M_{2}\bar
M_{3}}{2\bar M_{4}}+\frac{1}{8}\frac{(\bar M_{3})^{3}}{(\bar
M_{3})^{2}}\right)+
 \tilde M_{1}\left(\tilde \mu_{1}-\frac{1}{2}\tilde
 M_{1}\right)+\frac{\tilde M_{1}
\bar M_{3}}{\bar M_{4}},
\]
\[
h=\tilde \mu_{1}-P(0)\tilde M_{1}+\frac{\bar M_{3}}{\bar M_{4}}
\]
After the integration over $\chi_{\vec k}^{'}$ in (\ref{c4.19}) we get
\begin{equation}
\Xi=C\int \exp[E_{4}(\eta)](d\eta)^{N_{B}}, \label{c4.20}
\end{equation}
where
\begin{equation}
C=\Xi_{0}\Xi_{\xi}^{G}\Delta^{\xi}\Delta^{\eta}\exp({\cal
G}+a_{0}N_{B})\sqrt{2}^{N_{B}-1}, \label{c4.21}
\end{equation}
\begin{eqnarray}
E_{4}(\eta)&=&h\eta_{0}-\frac{1}{2\langle
N_{B}\rangle}\sum_{\vec k}d_{2}(k)\eta_{\vec k}\eta_{-\vec k}-\nonumber \\
&& \frac{a_{4}}{4!\langle N_{B}\rangle^{3}}\sum_{\vec k_{1}\ldots\vec
k_{4}}\eta_{\vec k_{1}}\eta_{\vec k_{2}}\eta_{\vec k_{3}}\eta_{\vec k_{4}}
\delta_{\vec k_{1}+\vec k_{2} +\vec
k_{3}+\vec k_{4}, \qquad   \vert\vec k_{i}\vert<B, \label{c4.22}}
\end{eqnarray}
\begin{equation}
a_{0}=\ln\Big[\frac{1}{\sqrt{\pi}}\Big(\frac{N_{B}}{N}\Big)^{1/4}\Big(\frac{3}{\vert
\tilde S_{4}\vert}\Big)^{1/4}\exp\Big(\frac{z^{2}}{4}\Big)U(0,z)\Big]
\label{c4.23}
\end{equation}
\begin{equation}
d_{2}(k)=a_{2}+P(k)   \label{c4.24}
\end{equation}
\begin{equation}
a_{2}=\sqrt{\frac{12}{\vert \tilde S_{4}\vert}\frac{\langle
N_{B}\rangle}{\langle N\rangle}}{\cal K}(z)  \label{c4.25}
\end{equation}
\begin{equation}
a_{4}=36\frac{\langle N_{B}\rangle}{\langle N\rangle}\frac{1}{\vert
\tilde S_{4}\vert}\Big[{\cal K}^{2}(z)+\frac{2}{3}{\cal
K}(z)-\frac{2}{3}\Big]
\label{c4.26}
\end{equation}
\begin{equation}
{\cal K}(z)= U(1,z)/U(0,z) \label{c4.27}
\end{equation}
\begin{equation}
z=\tilde S_{2}\sqrt{\frac{3}{\vert
\tilde S_{4}\vert}\frac{\langle N\rangle}{\langle N_{B}\rangle}}
\label{c4.28}
\end{equation}
\[
\tilde S_{n}=\sqrt{2}^{n}\tilde M_{n}/\langle N\rangle.
\]
Here $U(a,z)$ is the parabolic cylinder function.

As it is seen from (\ref{c4.22}), $E_{4}(\eta)$ has the form analogous
to the basic density measure of the $3D$ Ising model in an external field
\cite{yuk}. But the main difference is the dependence of
coefficients $a_{0}$, $a_{2}$ and $a_{4}$ (see
(\ref{c4.23})-(\ref{c4.28})) on the microscopic parameters of the system.

\section{Conclusions}

We propose the microscopic approach to the study of phase transitions and
critical phenomena in multicomponent mixtures. It is based on the CV method
with RS. This method allows us to take into account the short-range  and
long-range interactions  simultaneously.

We consider the task of the definition of the order parameter in a binary
mixture and show that it has a consistent and clear solution within the
framework of our approach.

After integration over CV $\xi_{\vec k}$ (which do not include the variable
connected with the order parameter) the basic density measure with respect
to CV $\eta_{\vec k}$  (Ginsburg-Landau--Wilson Hamiltonian) is constructed.
It is shown that the task can be reduced to the 3D Ising model.

\section*{Appendix A}
\setcounter{equation}{0}
A grand partition function of a two-component fluid system in  the
CV  representation with the RS can be written as \cite{patyuk3}:
\begin{displaymath}
\Xi=\Xi_{0}\Xi_{1},
\end{displaymath}
where
\begin{displaymath}
\Xi_{0}=\sum_{N_{a}=0}^{\infty}\sum_{N_{b}=0}^{\infty}\prod_{\gamma=a}^{b}
\exp\left[\frac{\beta\mu_{0}^{\gamma}N_{\gamma}}{N_{\gamma}!}\right]
\int(d\Gamma)\exp\left[-\frac{\beta}{2}\sum_{\gamma\delta}\sum_{ij}
\psi_{\gamma\delta}(r_{ij})\right].
\end{displaymath}
is the grand partition function of the RS; $\beta=\frac{1}{kT}$ is the
reciprocal temperature;
 $(d\Gamma)=\prod_{a,b}d\Gamma_{N_{\gamma}}$,
$d\Gamma_{N_{\gamma}}=d\vec r_{1}^{\gamma}d\vec r_{2}^{\gamma}\ldots d\vec
r_{N_{\gamma}}^{\gamma}$
is an element of the configurational space of the $\gamma$-th species;
$\mu_{0}^{\gamma}$ is the chemical potential of the $\gamma$-th
species in the RS.

The part of the grand partition function which is  defined  in
the CV phase space has the form of the functional integral:
\begin{equation}
\Xi_{1}=\int(d\rho)exp[\beta\sum_{\gamma}\mu_{1}^{\gamma}\rho_{0,\gamma}-
\frac{\beta}{2V}\sum_{\gamma\delta}\sum_{\vec
k}\tilde\phi_{\gamma\delta}(k) \rho_{\vec k,\gamma}\rho_{-\vec
k,\delta}]J(\rho_{a},\rho_{b}). \label{dA.1}
\end{equation}
Here,

1) $\mu_{1}^{\gamma}$ is a part of the chemical potential of the
$\gamma$-th species
\begin{displaymath}
\mu_{1}^{\gamma}= \mu_{\gamma}-\mu_{0}^{\gamma} +
\frac{\beta}{2v}\sum_{\vec k}\tilde \phi_{\gamma\gamma}(k)
\end{displaymath}
and is determined from the equation
\begin{equation}
\frac{\partial\ln\Xi_{1}}{\partial\beta\mu_{1}^{\gamma}} = \langle
 N_{\gamma}\rangle, \label{dA.2}
\end{equation}
$\mu_{\gamma}$ is the full chemical potential of the $\gamma$-th
species;

2)$\rho_{\vec k,\gamma}=\rho_{\vec k,\gamma}^{c}-i\rho_{\vec
k,\gamma}^{s}$ is the collective variable of the $\gamma$-th species,
the indices $c$ and $s$ denote the real part and the coefficient at
the imaginary part of $\rho_{\vec k,\gamma}$; $\rho_{\vec
k,\gamma}^{c}$ and $\rho_{\vec k,\gamma}^{s}$  describe the value of
$\vec k$-th fluctuation mode of the number of $\gamma$-th  species
particles. Each of $\rho_{\vec k,\gamma}^{c}$ and $\rho_{\vec
k,\gamma}^{s}$ takes all the real values from $-\infty$ to $+\infty$.
$(d\rho)$ is a volume element of the CV phase space:
\begin{displaymath}
(d\rho)=\prod_{\gamma}d\rho_{0,\gamma}{\prod_{\vec k\not=0}}'
d\rho_{\vec k,\gamma}^{c}d\rho_{\vec k,\gamma}^{s}.
\end{displaymath}
The prime means that the product over $\vec k$ is performed  in
the upper semispace;

3) $J(\rho_{a},\rho_{b})$ is the transition Jacobian to the
CV averaged on the RS:
\begin{eqnarray}
J(\rho_{a},\rho_{b}) & = & \int(d\nu)\prod_{\gamma=a}^{b}\exp\left
[i2\pi\sum_{\vec k} \nu_{\vec k,\gamma}\rho_{\vec k,\gamma}\right]
\exp\left [\sum_{n\geq 1}
\frac{(-i2\pi)^{n}}{n!}\times\right.\nonumber\\ &  & \left.
\sum_{\gamma_{1}\ldots\gamma_{n}} \sum_{\vec
k_{1}\ldots\vec k_{n}}
M_{\gamma_{1}\ldots\gamma_{n}}(\vec k_{1},\ldots,\vec k_{n}) \nu_{\vec
k_{1},\gamma_{1}}\ldots\nu_{\vec k_{n},\gamma_{n}}\right],
\label{dA.3}
\end{eqnarray}
where  variable $\nu_{\vec k,\gamma}$ is
conjugated to  CV $\rho_{\vec k,\gamma}$.
$M_{\gamma_{1}\ldots\gamma_{n}}(\vec k_{1}, \ldots,\vec k_{n})$ is the
$n$-th cumulant connected  with
$S_{\gamma_{1}\ldots\gamma_{n}}(k_{1},\ldots,k_{n})$, the $n$-particle
partial structure factor of the RS, by means of the relation
\begin{displaymath}
M_{\gamma_{1}\ldots\gamma_{n}}(\vec k_{1},\ldots,\vec
k_{n})= \sqrt[n]{N_{\gamma_{1}}\ldots
N_{\gamma_{n}}}S_{\gamma_{1}\ldots\gamma_{n}}
(k_{1},\ldots,k_{n})\delta_{\vec k_{1}+\cdots+\vec k_{n}},
\end{displaymath}
where $\delta_{\vec k_{1}+\cdots+\vec
k_{n}}$ is the Kroneker symbol.

In general, the dependence of
$M_{\gamma_{1}\ldots\gamma_{n}}(\vec k_{1}, \ldots,\vec k_{n})$ on
wave vectors $\vec k_{1},\ldots,\vec k_{n}$ is complicated. Hereafter we
shall replace $M_{\gamma_{1}\ldots\gamma_{n}}(\vec k_{1}, \ldots,\vec k_{n})$
by their values in long-wave length limit
$M_{\gamma_{1}\ldots\gamma_{n}}(0,\ldots,0)$;

4)  $\tilde \phi_{\gamma\delta}(k)$ is the Fourier transform of
attractive potential $\phi_{\gamma\delta}(r)$. Function
$\tilde \phi_{\gamma\delta}(k)$ satisfies the following requirements:
$\tilde \phi_{\gamma\delta}(k)$ is negative for small values of $\vec
k$ and $lim_{\vec k \to \infty}\tilde \phi_{\gamma\delta}(k)=0$. The
behaviour  of $\phi_{\gamma\delta}(r)$ in the region of the core
$r<\sigma_{\gamma\delta}$ must be determined from the conditions of
optimal separation of the interaction. For a very broad class of
potentials the general form of $\tilde \phi_{\gamma\delta}(k)$ is
presented in figure 6.
%%%%%%%%%%%%%%%%%%%%%%%%%%%%%%%%%%%%%%%%%%%%%%%%%%%%%%%%%%%%%%%%%%%
\begin{figure}[htbp]
\begin{center}
\epsfxsize 100mm
\epsfysize 100mm
\epsffile{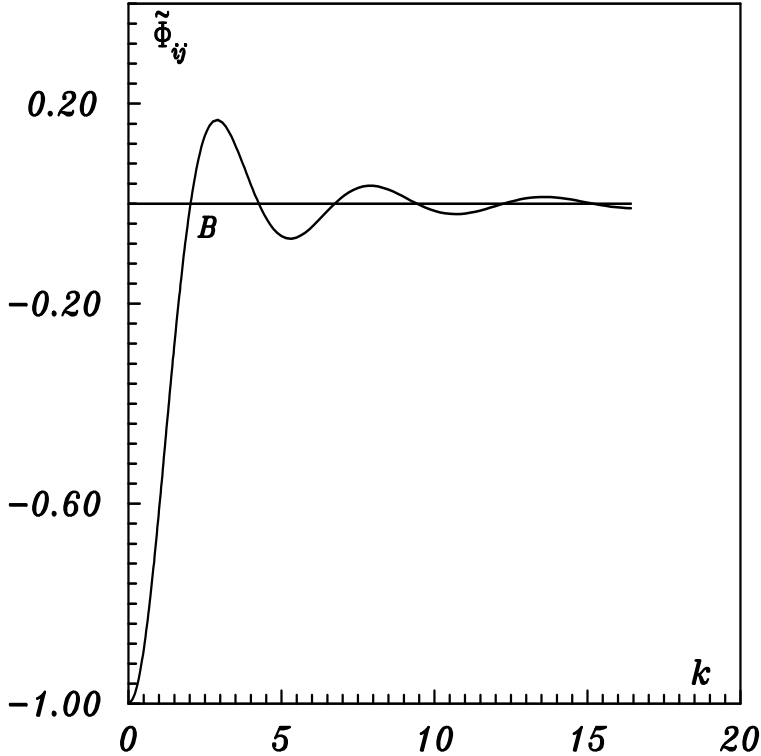}
\end{center}
\caption{The Fourier transform of the attractive potential
$\phi_{\gamma\delta}(r)$ }
\end{figure}
%%%%%%%%%%%%%%%%%%%%%%%%%%%%%%%%%%%%%%%%%%%%%%%%%%%%%%%%%%%%%%%%%%%%

We pass in (\ref{dA.1}) to CV $\rho_{\vec k}$ and $c_{\vec k}$
(according to $\omega_{\vec k}$ and $\gamma_{\vec k}$) by means of the
orthogonal linear transformation:
\begin{equation}
\rho_{\vec k} =  \frac{\sqrt{2}}{2}(\rho_{\vec k,a}+\rho_{\vec k,b})
\label{dA.4a}
\end{equation}
\begin{equation}
c_{\vec k} =  \frac{\sqrt{2}}{2}(\rho_{\vec k,a}-\rho_{\vec k,b}).
\label{dA.4b}
\end{equation}

As a result, for $\Xi_{1}$ we obtain formulas
(\ref{a2.1})-(\ref{a2.6}).

\section*{Appendix B}

The coefficients $A(k)$, $B(k)$, $C(k)$ and $D(k)$ have the forms:
\begin{eqnarray*}
A & = &\sqrt{2}\vert A_{12}\vert[4 A_{12}^{2} + (A_{11} -
A_{22})^{2}- (A_{11} - A_{22})\times \\
&  & \sqrt{(A_{11} - A_{22})^{2} + 4 A_{12}^{2}}]^{-1},\\
%%%%%%%%%%%%%%%%%%%%%%%%%%%%%%%%%%%%%%%%%%%%%%%%%%%%%%%%%%%%%%%
B & = &\sqrt{2}\vert A_{12}\vert[4 A_{12}^{2} + (A_{11} -
A_{22})^{2}+ (A_{11} - A_{22})\times \\
&  & \sqrt{(A_{11} - A_{22})^{2} + 4 A_{12}^{2}}]^{-1},\\
%%%%%%%%%%%%%%%%%%%%%%%%%%%%%%%%%%%%%%%%%%%%%%%%%%%%%%%%%%%%%%%%%%%%
C & = &-\frac{\sqrt{2}}{2}\frac{\vert A_{12}\vert}{(A_{12})}
[A_{11} - A_{22} - \sqrt{(A_{11} - A_{22})^{2} +
4A_{12}^{2}}]\times \\
&   &[4A_{12}^{2} + (A_{11} - A_{22})^{2}- (A_{11} - A_{22})
\sqrt{(A_{11} - A_{22})^{2} + 4A_{12}^{2}}]^{-1},\\
%%%%%%%%%%%%%%%%%%%%%%%%%%%%%%%%%%%%%%%%%%%%%%%%%%%%%%%%%%%%%%%%%%%%%
D & = &-\frac{\sqrt{2}}{2}\frac{\vert A_{12}\vert}{( A_{12})}
[A_{11} - A_{22} + \sqrt{(A_{11} - A_{22})^{2} +
4A_{12}^{2}}] \times \\
&  &[4A_{12}^{2} + (A_{11} - A_{22})^{2}+ (A_{11} - A_{22})
\sqrt{(A_{11} - A_{22})^{2} + 4A_{12}^{2}}]^{-1}.
\end{eqnarray*}
\section*{Appendix C}
The Helmholtz free energy of a binary mixture in the mean field
approximation can be written as
$$
f_{mf}=f_{id}+f_{ref}+f_{attr},
$$
where $f_{id}$ is the free energy of a binary mixture of ideal gases,
$f_{ref}$ is the free energy of a binary mixture of hard spheres
\cite{mansoori}:
\begin{eqnarray*}
f_{ref}=F_{ref}/\langle N\rangle
k_{B}T=-1.5(1-y_{1}+y_{2}+y_{3})+(3y_{2}+2y_{3})(1-\eta)^{-1}\\
+1.5(1-y_{1}-y_{2}-y_{3}/3)(1-\eta)^{-2}+(y_{3}-1)\ln(1-\eta),
\end{eqnarray*}
$$
y_{1}=\Delta_{12}\frac{1+\alpha}{\sqrt{\alpha}}, \qquad
y_{2}=\Delta_{12}\frac{\eta_{a}\alpha+\eta_{b}}{\sqrt{\alpha}\eta},
$$
$$
\Delta_{12}=\frac{\sqrt{\eta_{a}\eta_{b}}}{\eta}\frac{(\alpha-1)^{2}}{\alpha}
\sqrt{x(1-x)},
$$
$$
\eta_{a}=\frac{(1-x)\alpha^{3}\eta}{x+(1-x)\alpha^{3}}, \qquad
\eta_{b}=\frac{x\eta}{x+(1-x)\alpha^{3}}.
$$
$f_{attr}=F_{attr}/\langle N\rangle k_{B}T$ is the contribution due to
attraction between the particles:
$$
f_{attr}=-\frac{1}{2}\frac{\eta}{T^{*}(x+(1-x)\alpha^{3})}((1-x)^{2}+2x(1-x)r+
x^{2}q),
$$
where $T^{*}=k_{B}T\sigma^{3}\mid\tilde \phi_{aa}(0)\mid^{-1}\pi/6$ is the
dimensionless temperature.

\end{document}